\documentclass[aps,showpacs,superscriptaddress,twocolumn]{revtex4}%
\usepackage{amsfonts}
\usepackage{amsmath}
\usepackage{amssymb}
\usepackage{graphicx}%
\begin{document}
\title{Fractional quantum Hall effect of topological surface states under a strong
tilted magnetic field}
\author{Fawei Zheng }
\affiliation{LCP, Institute of Applied Physics and Computational Mathematics, P. O. Box
8009, Beijing 100088, China}
\author{Zhigang Wang}
\affiliation{LCP, Institute of Applied Physics and Computational Mathematics, P. O. Box
8009, Beijing 100088, China}
\author{Zhen-Guo Fu}
\affiliation{Beijing Computational Science Research Center, Beijing
100084, China}
\author{Ping Zhang}
\thanks{Corresponding author. Email address: zhang\_ping@iapcm.ac.cn}
\affiliation{LCP, Institute of Applied Physics and Computational Mathematics, P. O. Box
8009, Beijing 100088, China}
\affiliation{Beijing Computational Science Research Center, Beijing 100084, China}


\begin{abstract}
The fractional quantum Hall effect (FQHE) of topological surface-state
particles under a tilted strong magnetic field is theoretically studied by
using the exact diagonalization method. The Haldane's pseudopotentials for the
Coulomb interaction are analytically obtained. The results show that by
increasing the in-plane component of the tilted magnetic field, the FQHE state
at $n$=$0$ Landau level (LL) becomes more stable, while the stabilities of
$n$=$\pm1$ LLs become weaker. Moreover, we find that the excitation gaps of
the $\nu=1/3$ FQHE states increase as the tilt angle is increased.

\end{abstract}
\maketitle

The fractional quantum Hall effect (FQHE) in conventional tow-dimensional
electron gas (2DEG) has been studied intensively in the last 30 years
\cite{Laughlin, Halperin, Haldane,Girvin1990,Das1997,Murthy,Jain1989} because
of its rich physical properties. Recently, the unconventional sequence of FQHE
in graphene, where the valley isospin combined with the usual electron spin
yields fourfold degenerate Landau levels (LLs), has been observed
\cite{Feldman,Bolotin,Du,Dial,Dean1}. Differing from the case of the
conventional 2DEG, due to the Dirac nature of the electrons and the additional
symmetry, the FQHE is modified and new incompressible ground states should be
conjectured in graphene \cite{Toke, Apalkov, Goerbig,Papic,Ghaemi}. Moreover,
the linear energy dispersion also modifies the inter-electron interactions,
which implies a specific dependence of the ground state energy and energy gap
on LL index \cite{Toke, Apalkov, Goerbig}. The graphene-like integer quantum
Hall effect has been observed in strained bulk HgTe \cite{Brune}\ and the LL
spectrum has also been measured in three-dimensional topological insulator
(TI) material Bi$_{2}$Se$_{3}$ by using scanning tunneling microscopy
\cite{Cheng, Hanaguri}. In addition, features in the Hall resistance at
fractional filling factors have been speculated to be related to the FQHE of
TIs \cite{Analytis, Xiong,Apalkov2}. However, the strength of the FQHE will be
different from that in graphene because the LLs of the two surfaces of the TI
thin film can mix with each other \cite{Apalkov2}.

Furthermore, previous studies suggest that when the spin freedom and the
parallel magnetic field are taken into account, the spin-reversed
quasiparticle (quasihole) excitations could be observed in certain FQHE gaps
of GaAs-based quantum wells \cite{Chakraborty,Furneaux}. The
quasiparticle-quasihole energy gap increases with increasing the tilt angle of
magnetic field, and the electron-hole symmetry are broken
\cite{Haug,Chakraborty1} due to the subband-LL coupling in FQHE. The
contrasting behavior of the higher filling ($\nu=5/2$ and $\nu=7/3$) FQHE
states and the relevance of a skyrmion spin texture at $\nu=5/2$ associated
with small Zeeman energy in wide GaAs/AlGaAs quantum wells under a tilted
magnetic field have been studied in recent experiments \cite{Dean,Liu2012}.
Surprisingly, with increasing the tilt angle the FQHE states with
even-denominator filling factors may transform to the compressible Fermi
liquid state from the incompressible state.

\begin{figure}[ptb]
\begin{center}
\includegraphics[width=1.\linewidth]{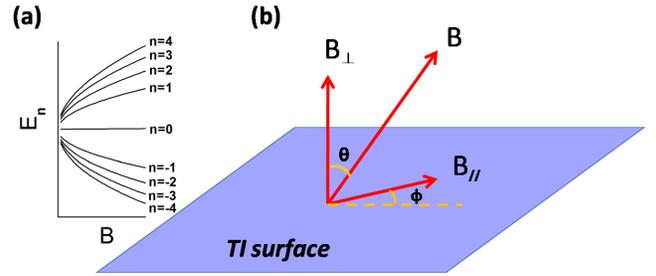}
\end{center}
\caption{(Color online) (a) The fermion Landau levels on the topological
insulator surface under an external magnetic field. The corresponding
experimental set sketch is shown in (b).}%
\end{figure}

Unambiguous characterizations of the FQHE of Dirac fermions on TI surface (or
graphene) in the presence of a tilted magnetic field, however, is still
missing in literature. Owing to the unique spin chirality of Dirac fermions
induced by intrinsic strong spin-orbit coupling in TI materials, the effect of
tilted magnetic field on the FQHE on TI surface should remarkably differ from
that in conventional 2DEG. Therefore, because of its importance both from
basic point of interest and to the analysis of unconventional properties of
TI-based FQHE, in the present paper we address the stability of the FQHE on TI
under a strong tilted magnetic field by presenting an attempt at the
theoretical evaluation of the effective pseudopotentials of the
electron-electron interactions and the ground (excitation) state nature at
$1/3$\texttt{-}FQHE at lowest LLs problems. We will show that by increasing
the in-plane component of the tilted magnetic field, the FQHE state at $n$=$0$
LL becomes more stable, while the situation for the $n$=$\pm1$ LLs just have
an opposite trend, becoming weaker. Besides, we find that the excitation gaps
of the $\nu=1/3$ FQHE states increase as the tilt angle is increased.

The single particle low-energy effective Hamiltonian of TI surface in the
presence of a tilted magnetic field, $\vec{B}=\vec{B}_{\bot}+\vec
{B}_{\parallel}=B(\hat{x}\sin\theta\cos\phi+\hat{y}\sin\theta\sin\phi+\hat
{z}\cos\theta)$, where $\theta$ is the tilt angle and $\varphi$ is the
azimuthal angle with respect to two-dimensional (2D) plane, is expressed as%
\begin{equation}
H_{0}=v_{F}\vec{\sigma}\cdot\vec{\Pi}+g\mu_{B}\vec{\sigma}\cdot\vec
{B},\label{1}%
\end{equation}
where $v_{F}$ $=3\times10^{5}$ m/s is the Fermi velocity and $\vec{\Pi}%
=\vec{p}+e\vec{A}/c$ is the 2D canonical momentum. We choose the symmetric
gauge for the vector potential of vertical magnetic field $\vec{A}_{\bot
}=B_{\bot}(-y/2,x/2,0)$, and the Landau gauge for the vector potential of
in-plane magnetic potential $\vec{A}_{\parallel}=(zB_{y},-zB_{x},0)$. Because
the motions of electrons are confined in $z=0$ plane, the electron canonical
momentum is only relevant to $\vec{A}_{\bot}$. After introducing the ladder
operators $a^{\dag}=\frac{1}{\sqrt{2}}\left(  \frac{\kappa}{2l}-2l\partial
_{\bar{\kappa}}\right)  $, $a=\frac{1}{\sqrt{2}}\left(  \frac{\bar{\kappa}%
}{2l}+2l\partial_{\kappa}\right)  $ with $\kappa=x+iy$ and the magnetic length
$l=\sqrt{\frac{\hbar c}{eB_{\bot}}}$, we can rewrite the Hamiltonian (\ref{1})
as
\begin{equation}
H_{0}\text{=}\left(
\begin{array}
[c]{cc}%
g\mu_{B}B_{\bot} & -i\frac{\sqrt{2}\hbar v_{F}}{l}a\text{+}g\mu_{B}%
B_{\shortparallel}e^{-i\phi}\\
i\frac{\sqrt{2}\hbar v_{F}}{l}a^{\dag}\text{+}g\mu_{B}B_{\shortparallel
}e^{i\phi} & -g\mu_{B}B_{\bot}%
\end{array}
\right)  .\label{2}%
\end{equation}
The isotropic property of electrons in 2D plane requires that the physical
results should be independent of the azimuth angle $\phi$. One can see in the
following text that the Haldane's pseudopotentials rightly satisfy this
symmetry condition. By employing the perturbation method and only keeping the
first-order approximation of the wave function, the eigenstates of Hamiltonian
(\ref{2}) may be written as
\begin{align}
\Psi_{n,m} &  =\left(
\begin{array}
[c]{c}%
\Psi_{n,m}^{1}\\
\Psi_{n,m}^{2}%
\end{array}
\right)  \nonumber\\
&  =\left(
\begin{array}
[c]{c}%
\alpha_{n}\phi_{|n|-1,m}+A_{n}^{1}\phi_{|n|,m}+A_{n}^{2}\phi_{|n|-2,m}\\
\beta_{n}\phi_{|n|,m}+B_{n}^{1}\phi_{|n|-1,m}+B_{n}^{2}\phi_{|n|+1,m}%
\end{array}
\right)  ,
\end{align}
where $\phi_{nm}$ is the eigenstate of the 2D Hamiltonian with
non-relativistic quadratic dispersion relation in $n$th LL with angular
momentum $m$. Here,
\begin{align}
\alpha_{n} &  =\left\{
\begin{array}
[c]{ll}%
0, & \hbox{n=0}\\
\frac{-i\text{sgn}(n)\cos\vartheta_{n}}{\sqrt{2\left(  1-\text{sgn}%
(n)\sin\vartheta_{n}\right)  }}, & \hbox{n$\neq$0}
\end{array}
\right.  ,\nonumber\\
\beta_{n} &  =\left\{
\begin{array}
[c]{ll}%
1, & \hbox{n=0}\\
\sqrt{\left(  1-\text{sgn}(n)\sin\vartheta_{n}\right)  /2}, & \hbox{n$\neq$0}
\end{array}
\right.  ,
\end{align}
where $\vartheta_{n\neq0}=$tan$^{-1}\frac{g\mu_{B}B_{\bot}}{\sqrt{2|n|}\hbar
v_{F}l^{-1}}$. The other coefficients are expressed as $A_{n}^{1}%
=\frac{\left(  \varepsilon_{n}+b\right)  d\beta_{n}}{\varepsilon_{n}^{2}%
-b^{2}-c^{2}(|n|+1)}$, $A_{n}^{2}=\frac{icd^{\ast}\alpha_{n}\sqrt{|n|-1}%
}{\varepsilon_{n}^{2}-b^{2}-c^{2}(|n|-1)}$, $B_{n}^{1}=\frac{\left(
\varepsilon_{n}-b\right)  d^{\ast}\alpha_{n}}{\varepsilon_{n}^{2}-b^{2}%
-c^{2}(|n|-1)}$, and $B_{n}^{2}=\frac{icd\beta_{n}\sqrt{|n|+1}}{\varepsilon
_{n}^{2}-b^{2}-c^{2}(|n|+1)}$ with $b=g\mu_{B}B_{\bot}$, $d=g\mu
_{B}B_{\shortparallel}e^{-i\phi}$, $c=\sqrt{2}\hbar v_{F}l^{-1}$. The
corresponding eigenvalues are given by $\varepsilon_{0}=-g\mu_{B}B_{\bot}=-b$,
and $\varepsilon_{n\neq0}=$sgn$(n)\sqrt{\left(  g\mu_{B}B_{\bot}\right)
^{2}+|\sqrt{2|n|}\hbar v_{F}l^{-1}|^{2}}$. In the limit of $B_{\shortparallel
}$=$0$, the Hamiltonian (2) can be analytically solved
\cite{Wang1,Wang,DaSilva}.

In order to investigate the properties of TI surface FQHE, we first consider
Coulomb interaction $V(\mathbf{r})$ $=e^{2}/\epsilon r$ between two electrons
in $n$th LL with relative angular momentum $m$. The Haldane's pseudopotential
is given by
\begin{equation}
V_{\text{eff}}^{(n,m)}\mathtt{=}\frac{1}{2}%
{\displaystyle\sum\limits_{\mathbf{q}}}
\frac{2\pi e^{2}}{\epsilon q}\left[  \mathcal{F}_{n}(q)\right]  ^{2}%
L_{m}\left(  q^{2}l^{2}\right)  e^{-\frac{q^{2}l^{2}}{2}},\label{e5}%
\end{equation}
where $L_{m}(x)$ are the Lagueree polynomials. The form factor $\mathcal{F}%
_{n}(q)$ in Eq. (\ref{e5}) is given by%
\begin{equation}
\mathcal{F}_{n}(q)=\langle\Psi_{n,m}^{1}|e^{-i\mathbf{q}\cdot\mathbf{\eta}%
}|\Psi_{n,m}^{1}\rangle+\langle\Psi_{n,m}^{2}|e^{-i\mathbf{q}\cdot
\mathbf{\eta}}|\Psi_{n,m}^{2}\rangle,\label{e6}%
\end{equation}
where $\mathbf{\eta}$=$\mathbf{r}-\mathbf{R}$ is the cyclotron variable with
guiding-center $\mathbf{R}$. The explicit expression of $\mathcal{F}_{n}(q)$
is derived to be written as \begin{widetext}%
\begin{align}
\mathcal{F}_{n}(q) &  =\left[  \left(  |\beta_{n}|^{2}+|A_{n}^{1}|^{2}\right)
L_{|n|}\left(  |s|^{2}\right)  +|B_{n}^{2}|^{2}L_{|n|+1}\left(  |s|^{2}%
\right)  \right.  \nonumber\\
&  +\left(  |\alpha_{n}|^{2}+|B_{n}^{1}|^{2}\right)  L_{|n|-1}\left(
|s|^{2}\right)  +|A_{n}^{2}|^{2}L_{|n|-2}\left(  |s|^{2}\right)  \nonumber\\
&  \left.  +\alpha_{n}^{\ast}A_{n}^{2}X_{|n|-1,|n|-2}(s)+\alpha_{n}%
A_{n}^{2\ast}X_{|n|-2,|n|-1}(s)+A_{n}^{1}A_{n}^{2\ast}X_{|n|-2,|n|}%
(s)+A_{n}^{1\ast}A_{n}^{2}X_{|n|,|n|-2}(s)\right.  \nonumber\\
&  \left.  +\beta_{n}^{\ast}B_{n}^{2}X_{|n|,|n|+1}(s)+\beta_{n}B_{n}^{2\ast
}X_{|n|+1,|n|}(s)+B_{n}^{2}B_{n}^{1\ast}X_{|n|-1,|n|+1}(s)+B_{n}^{2\ast}%
B_{n}^{1}X_{|n|+1,|n|-1}(s)\right.  \nonumber\\
&  \left.  +\left(  \alpha_{n}^{\ast}A_{n}^{1}+\beta_{n}B_{n}^{1\ast}\right)
X_{|n|-1,|n|}(s)+\left(  \alpha_{n}A_{n}^{1\ast}+\beta_{n}^{\ast}B_{n}%
^{1}\right)  X_{|n|,|n|-1}(s)\right]  e^{-\frac{|s|^{2}}{2}},\label{e7}%
\end{align}
where $X_{n,n^{\prime}}(s)$=$X_{n^{\prime},n}^{\ast}(-s)=(-is)^{n^{\prime}%
-n}\sum_{m=0}^{\text{min}(n,n^{\prime})}\frac{(-|s|^{2})^{n-m}\sqrt
{n!n^{\prime}!}}{(n-m)!(n^{\prime}-m)!m!}$ with $s\equiv l(q_{x}+iq_{y}%
)/\sqrt{2}$.
\end{widetext}

\begin{figure}[ptb]
\begin{center}
\includegraphics[width=1.\linewidth]{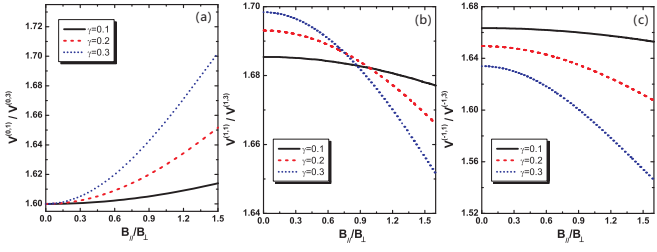}
\end{center}
\caption{(Color online) The ratio of the first and third relative angular
momentum pseudopotentials for the Coulomb interaction in the $n$=$0$ LL (a),
$n$=$1$ LL (b), and $n$=$-1$ LL (c) as a function of $B_{\shortparallel
}/B_{\bot}$. The black sold, red dashed, and blue dotted lines correspond to
$\gamma\equiv$ $g\mu_{B}B_{\bot}/(\hbar v_{F}l^{-1})$=$0.1$, $0.2$, $0.3$,
respectively. }%
\end{figure}

A typical way to predict the stability of the FQHE is to consider only
$m\mathtt{=}1$ and $m\mathtt{=}3$ pseudopotentials, since $V_{\text{eff}%
}^{(n,m)}$ falls off quickly as $m$ increases. The FQHE state may be observed
when the ratio between $m\mathtt{=}1$ and $m\mathtt{=}3$ pseudopotentials is
larger than $1.3\mathtt{-}1.5$, because the composite fermion wavefunction is
accurate when the interaction is short range, and remains a good approximation
when $V_{\text{eff}}^{(n,1)}/V_{\text{eff}}^{(n,3)}$ is large enough
\cite{Jain1989}. For the purpose of studying the stability of the FQHE states
under a tilted magnetic field, we also employ this crucial criterion and
choose $1.5$ as a critical value. In other words, when $V_{\text{eff}}%
^{(n,1)}/V_{\text{eff}}^{(n,3)}$ is smaller than $1.5$, we conclude the FQHE
state can not be observed.\begin{figure}[ptb]
\begin{center}
\includegraphics[width=1.\linewidth]{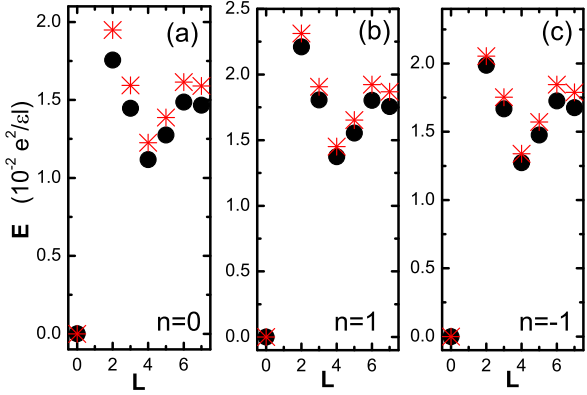}
\end{center}
\caption{(Color online) Exact energy per electron versus the angular momentum
$L$ for $N=7$ electrons at $\nu=1/3$ FQHE states for (a) $n=0$, (b) $n=1$ and
(c) $n=-1$ LLs. We choose the parameter $g\mu_{B}B_{\bot}/(\hbar v_{F}l^{-1}%
)$=$0.2$. The dots and stars correspond to $B_{\shortparallel}$=$0$ and
$B_{\shortparallel}$=$B_{\bot}$ respectively.}%
\end{figure}

For simplicity, let us first consider the stability of the FQHE state at $n=0$
LL, whose form factors is $\mathcal{F}_{0}(q)$=$\left[  L_{0}\left(
|s|^{2}\right)  \mathtt{+}\frac{|d|^{2}}{c^{2}}L_{1}\left(  |s|^{2}\right)
\mathtt{+}\frac{d^{\ast}s^{\ast}-ds}{c}\right]  e^{-\frac{|s|^{2}}{2}}$.
Substituting this $\mathcal{F}_{0}(q)$ into Eq. (\ref{e5}), one easily obtain
the Haldane's pseudopotential for the $0$-th LL,
\begin{align}
V_{\text{eff}}^{(0,m)} &  =\frac{1}{2}%
{\displaystyle\sum\limits_{\mathbf{q}}}
\frac{2\pi e^{2}}{\epsilon q}L_{m}\left(  2|s|^{2}\right)  e^{-2|s|^{2}}\\
&  \times\left[  1+2\left\vert \frac{d}{c}\right\vert ^{2}\left(
1-2|s|^{2}\right)  +\left\vert \frac{d}{c}\right\vert ^{4}\left(
1-|s|^{2}\right)  \right]  .\nonumber
\end{align}
One can find that, differing from the case under a fully perpendicular
magnetic field, the form factor in a tilted magnetic filed case is dependent
on the in-plane component of the external magnetic field. The numerical result
of $V_{\text{eff}}^{(0,1)}/V_{\text{eff}}^{(0,3)}$ as a function of in-plane
component of the external magnetic field $B_{\shortparallel}$ is shown in Fig.
2(a), where different curves correspond to different given perpendicular
components of the external magnetic field $B_{\bot}$. From Fig. 2(a) one can
see that $V_{\text{eff}}^{(0,1)}/V_{\text{eff}}^{(0,3)}$ always increases with
increasing $B_{\shortparallel}$. Based on this fact and on that the minimum
value of $V_{\text{eff}}^{(0,1)}/V_{\text{eff}}^{(0,3)}$ is $\mathtt{\sim}1.6$
at $B_{\shortparallel}$=$0$ (larger than the critical value $1.5$), we can
immediately conclude that the FQHE state at $n=0$ LL is more stable by
increasing the in-plane component of the tilted magnetic field.

Now we turn out to study the cases for the $n$=$\pm1$ LLs, whose
pseudopotentials can be written as \begin{widetext}%
\begin{align}
V_{\text{eff}}^{(n=\pm1,m)} &  =\frac{1}{2}%
{\displaystyle\sum\limits_{\mathbf{q}}}
\frac{2\pi e^{2}}{\epsilon q}L_{m}\left(  2|s|^{2}\right)  e^{-2|s|^{2}%
}  \times\left[  \left(  \left\vert \alpha_{n}^{\text{new}}\right\vert
^{2}L_{0}(|s|^{2})+\left\vert \beta_{n}^{\text{new}}\right\vert ^{2}%
L_{1}(|s|^{2})+2\left\vert \beta_{n}\right\vert ^{2}\left\vert \frac{d}%
{c}\right\vert ^{2}L_{2}(|s|^{2})\right)  ^{2}\right.  \nonumber\\
&  -8\left\vert \alpha_{n}\beta_{n}\right\vert ^{2}|s|^{2}\left\vert \frac
{d}{c}\right\vert ^{2}\frac{b^{2}}{c^{2}}+\frac{1}{2}\left\vert \alpha
_{n}\beta_{n}\right\vert ^{2}|s|^{4}\left\vert \frac{\varepsilon_{n}-b}%
{c}\right\vert ^{2}\left\vert \frac{d}{c}\right\vert ^{4} \left.  -2\left\vert \beta_{n}\right\vert ^{4}|s|^{2}\left\vert \frac{d}%
{c}\right\vert ^{2}\left(  1-\frac{|s|^{2}}{2}\right)  \right]  ,\label{e9}
\end{align}
\end{widetext}where $\left\vert \alpha_{n}^{\text{new}}\right\vert
^{2}=\left\vert \alpha_{n}\right\vert ^{2}\left(  1+\left\vert \frac
{\varepsilon_{n}-b}{c}\right\vert ^{2}\left\vert \frac{d}{c}\right\vert
^{2}\right)  $, and $\left\vert \beta_{n}^{\text{new}}\right\vert
^{2}=\left\vert \beta_{n}\right\vert ^{2}\left(  1+\left\vert \frac
{\varepsilon_{n}+b}{c}\right\vert ^{2}\left\vert \frac{d}{c}\right\vert
^{2}\right)  $. Figures 2(b) and 2(c) respectively plot $V_{\text{eff}%
}^{(n=\pm1,1)}/V_{\text{eff}}^{(n=\pm1,3)}$ versus $B_{\shortparallel}$, from
which one can find that $V_{\text{eff}}^{(n=\pm1,1)}/V_{\text{eff}}%
^{(n=\pm1,3)}$ always decrease with increasing $B_{\shortparallel}$. These
results imply that the Haldane's pseudopotentials for the $n$=$\pm1$ LLs the
FQHE states turn to become instable by increasing the in-plane component of
the tilted magnetic field, which is quite different from the case for $n$=$0$ LL.

\begin{figure}[ptb]
\begin{center}
\includegraphics[width=1.\linewidth]{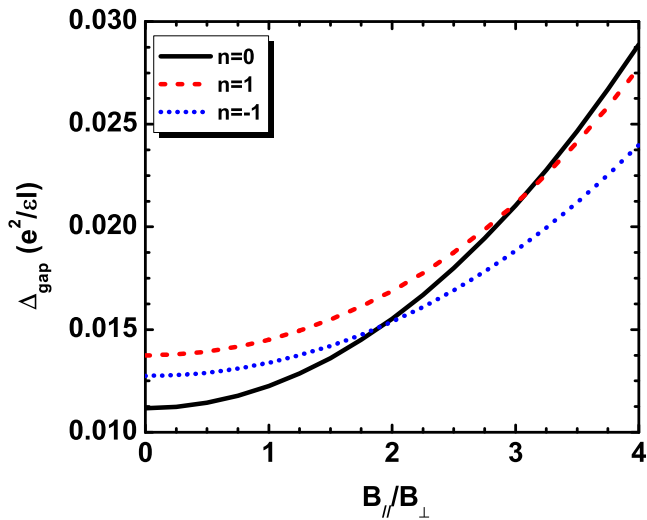}
\end{center}
\caption{(Color online) The gap width between the ground state and the first
excited state at $\nu=1/3$ FQHE states for LLs $n=0$ and $\pm1$ as a function
of $B_{//}/B_{\bot}$. The other parameters are $N=7$ and $g\mu_{B}B_{\bot
}/(\hbar v_{F}l^{-1})$=$0.2$. }%
\end{figure}We also investigate the energy spectra of the many-body states at
fractional filling $\nu=1/3$ of the LLs $n$=$0$ and $\pm1$ under a tilted
magnetic field by numerically diagonalizing the many-body Hamiltonian in the
spherical geometry. Figure 3 shows the energy spectra for $N=7$ electrons at
$\nu=1/3$ FQHE states for $n$=$0$ and $1$ LLs with different in-plane
components of the external magnetic field, $B_{||}$=$0$ (black dots) and
$B_{||}$=$B_{\bot}$ (red stars). One can clearly see that the excitation gap
width at $B_{||}$=$B_{\bot}$ is larger than that at $B_{||}$=$0$. To more
clearly see how the gap width changes with the tilt angle, in Fig. 4 we
exhibit the gap width between the ground state and the lowest excited state as
a function of the in-plane component of the magnetic field. It is obvious that
by increasing the in-plane component of the magnetic field, the gap widths at
$\nu=1/3$ FQHE states for LLs $n$=$0$ and $\pm1$ become larger and larger,
which is similar to the conventional 2DEG cases \cite{Chakraborty1989,Halonen}%
.

In summary, we theoretically investigated the FQHE in TIs under a tilted
strong magnetic field. The single particle wave function was obtained by using
a simple perturbation method. The effective pseudopotentials of the
electron-electron interactions and the ground (excited) state energy spectra
for $1/3$\texttt{-}FQHE at lowest LLs were calculated within the exact
diagonalization approach. We have shown that in the presence of a in-plane
component of the tilted magnetic field, the FQHE state at $n$=$0$ LL becomes
more stable, while the stabilities of $n$=$\pm1$ LLs become weaker. Moreover,
we have also found that the excitation gaps of the $\nu=1/3$ FQHE states
increase as the tilt angle is increased.

This work was supported by Natural Science Foundation of China under Grants
No. 11274049, No. 90921003, and No. 11004013, and by the National Basic
Research Program of China (973 Program) under Grant No. 2009CB929103.

\end{document}